\documentclass[useAMS,usenatbib,aas_macros]{mn2e}

\usepackage{epsfig}
\usepackage{amssymb}
\usepackage{natbib}
\usepackage{aas_macros}

\begin{document}

\def\PapI{Shankar et al. (2009c)}
\def\PapInp{Shankar et al. 2009c}
\newcommand{\CIV}{C~{\sc iv}}
\def\sarc{$^{\prime\prime}\!\!.$}
\def\arcsec{$^{\prime\prime}\, $}
\def\arcmin{$^{\prime}$}
\def\kms{${\rm km\, s^{-1}}$}
\def\degr{$^{\circ}$}
\def\re{$R_e$}
\def\rb{$R$}
\def\phire{$\Phi(R_e)$}
\def\phirez{$\Phi(R_e,z)$}
\def\phiv{$\Phi(\sigma)$}
\def\sis{$\sigma$}
\def\seco{$^{\rm s}\!\!.$}
\def\ls{\lower 2pt \hbox{$\;\scriptscriptstyle \buildrel<\over\sim\;$}}
\def\gs{\lower 2pt \hbox{$\;\scriptscriptstyle \buildrel>\over\sim\;$}}
\def\mbh{$M_{\rm BH}$}
\def\mstar{$M_{\rm STAR}$}
\def\msun{${\rm M_{\odot}}$}
\def\MMs{$M_{\rm dyn}/M_{\rm STAR}$}
\def\ML{$M_{\rm dyn}/L_r$}
\def\MsL{$M_{\rm STAR}/L_r$}
\def\MLc{$M_{\rm dyn}/L_{\rm corr}$}
\def\MdM{$M_{\rm dyn}/M_{\rm STAR}$}
\def\Lcorr{$L_r^{\rm corr}$}
\def\lsun{${\rm L_{\odot}}$}

\newcommand{\kpch}{$h^{-1}\,\mbox{kpc}$\,}

\title{Sizes and ages of SDSS ellipticals:
Comparison with hierarchical galaxy formation models}

\author[F. Shankar et al.]
{Francesco Shankar$^{1}$\thanks{E-mail:$\;$shankar@mpa-garching.mpg.de},
Federico Marulli$^{2}$, Mariangela Bernardi$^{3}$,
\newauthor
Xinyu Dai$^{4}$, Joseph B. Hyde$^{3}$ and Ravi K. Sheth$^{3}$
\\
$1$ Max-Planck-Instit\"{u}t f\"{u}r Astrophysik,
Karl-Schwarzschild-Str. 1, D-85748, Garching, Germany\\
$2$ Dipartimento di Astronomia, Universit\'{a} degli
Studi di Bologna, via Ranzani 1, I-40127 Bologna, Italy\\
$3$ Department of Physics and Astronomy, University of Pennsylvania,
209 South 33rd St, Philadelphia, PA 19104\\
$4$ Astronomy Department, University of Michigan, Ann Arbor, MI, 48109}

\date{}
\pagerange{\pageref{firstpage}--\pageref{lastpage}} \pubyear{2008}
\maketitle

\label{firstpage}

\begin{abstract}
In a sample of about 45,700 early-type galaxies extracted from
SDSS, we find that the shape, normalization, and dispersion around
the mean size-stellar mass relation is the same for young and old
systems, provided the stellar mass is greater than
$3\times 10^{10}\, $\msun.
This is difficult to reproduce in pure passive evolution models,
which generically predict older galaxies to be much more compact than
younger ones of the same stellar mass.  However, this aspect of our
measurements is well reproduced by hierarchical models of galaxy
formation.  Whereas the models predict more compact galaxies at
high redshifts, subsequent minor, dry mergers increase the sizes
of the more massive objects, resulting in a flat size-age relation
at the present time.  At lower masses, the models predict that
mergers are less frequent, so that the expected anti-correlation
between age and size is not completely erased.  This is in good
agreement with our data:  below $3\times 10^{10}\, $\msun, the
effective radius \re\ is a factor of $\sim 2$ lower for older
galaxies.  These successes of the models are offset by the fact
that the predicted sizes have other serious problems, which we
discuss.
\end{abstract}

\begin{keywords}
galaxies: structure -- galaxies: formation -- galaxies: evolution --
cosmology: theory
\end{keywords}

\section{Introduction}
\label{sec|intro}

According to the standard cosmological paradigm of structure
formation and evolution, dark matter (DM) halos have grown
hierarchically, through the merging together of smaller units into
ever larger systems \citep{Press74,Bond91,Lacey93,ShethMF}.
In this scenario, galaxies form inside this hierarchically growing
system of DM halos \citep{White79,White91}.
Semi-analytical models (SAMs) of galaxy formation
\citep[e.g.,][]{Cole00,Benson03,Granato04,Menci04,Bower06,Cattaneo06,Croton06,DeLucia06,KochfarSilk06Rez,KochfarSilk06origin,Hopkins06,Monaco07}
have now been able to reproduce several properties of the local
galaxy population. In particular, the local galaxy luminosity
function has been successfully matched both at the high and low
luminosity ends owing to the implementation of models for the feedback
from both stellar evolution and active galactic nuclei (AGN)
\citep[e.g.,][]{Benson03,Granato04,Granato06,DiMatteo05,Bower06,Cattaneo06,Croton06,DeLucia06,Menci06,Monaco07}.
While this feedback significantly reduces the amount of cooling baryons
in the host halos, it also seems to be a promising tool to account
for the AGN luminosity functions and mean trends in the stellar-halo
mass relations
\citep[e.g.,][]{WL03,Scannapieco04,Ciras05,DiMatteo05,Sazonov05,Vittorini05,Hopkins06,Lapi06,Shankar06,Benson07,Fontanot07,Malbon07,Marulli08,ShankarCrocce}.

In addition to the luminosities, masses, and metallicities of galaxies,
their \emph{sizes}, measured at low and high redshift, provide strong
constraints on galaxy formation models \citep[e.g.,][]{Ciras05,KochfarSilk06Rez,Almeida07,Fan08,GonzalezSAM08,Hop08FP,Bernardi09,vanderwel09}.
Recent observations show that galaxies of a given stellar
mass are more compact \citep[e.g.,][]{Trujillo06,Trujillo07,Cimatti08,
Buitrago08b,Chapman08,Franx08,Saracco08,Tacconi08,Vanderwel08,Vandokkum08,Younger08,Damjanov09,Williams09},
and, at least some, have higher velocity dispersions \citep[e.g.,][]{Cenarro09,vandokkum09}
at redshifts greater than unity; this is also true of BCGs, even for small lookback times \citep{Bernardi09}.
\citet{Saracco08} also showed that at $z\sim 1.4$,
while older galaxies at fixed stellar mass tend
to lie a factor of $\sim 2-3$ below
the size-stellar mass relation characterizing
local early-type galaxies (\citealt{Shen03}), younger galaxies
are consistent with it.
After analyzing a sample of 12 very massive
galaxies at similar redshift, \citet{Mancini09} suggested a downsizing scenario
in sizes, with the most massive galaxies approaching the local size-mass
relation earlier then less massive ones. \citet{Cappellari09}
also discussed that two
massive galaxies with individual spectra at $1.4\lesssim z \lesssim 2$
are similar to local counterparts, and another 7 galaxies
with velocity dispersion from staked spectrum, are consistent
to the most dense local galaxies of the same mass.

Small sizes (and, at fixed mass, higher velocity
dispersions) at high redshifts are not unexpected:
they result if galaxies at higher redshifts formed through more
gas-rich and dissipative mergers \citep[e.g.,][]{Robertson06}.
Moreover, the gas fractions and the overall
density of the universe decrease with time \citep[e.g.,][and references therein]{Hop08FP}
implying, on average, less dense remnants at later times.
Therefore if galaxies continuously
form at different epochs, it is expected
that in deep observation at $z>1.5-2$, galaxies
at fixed stellar mass might
follow a distribution of sizes, with older galaxies born
from gas-richer events, being
more compact.

However, the problem is to explain how all galaxies,
irrespective of their age, evolve in time on the same
local size-mass relation.
As shown by \citet{Vandokkum08}, \citet{ShankarBernardi},
and further discussed here, a simple pure monolithic collapse
followed by strictly passive evolution is not satisfactory, as this
does not explain why the extremely small sizes and high densities
of massive galaxies at high-$z$ have no local counterparts in SDSS
\citep[also see][]{Trujillo09}.

\begin{figure*}%[ht!]
\includegraphics[width=17.5truecm]{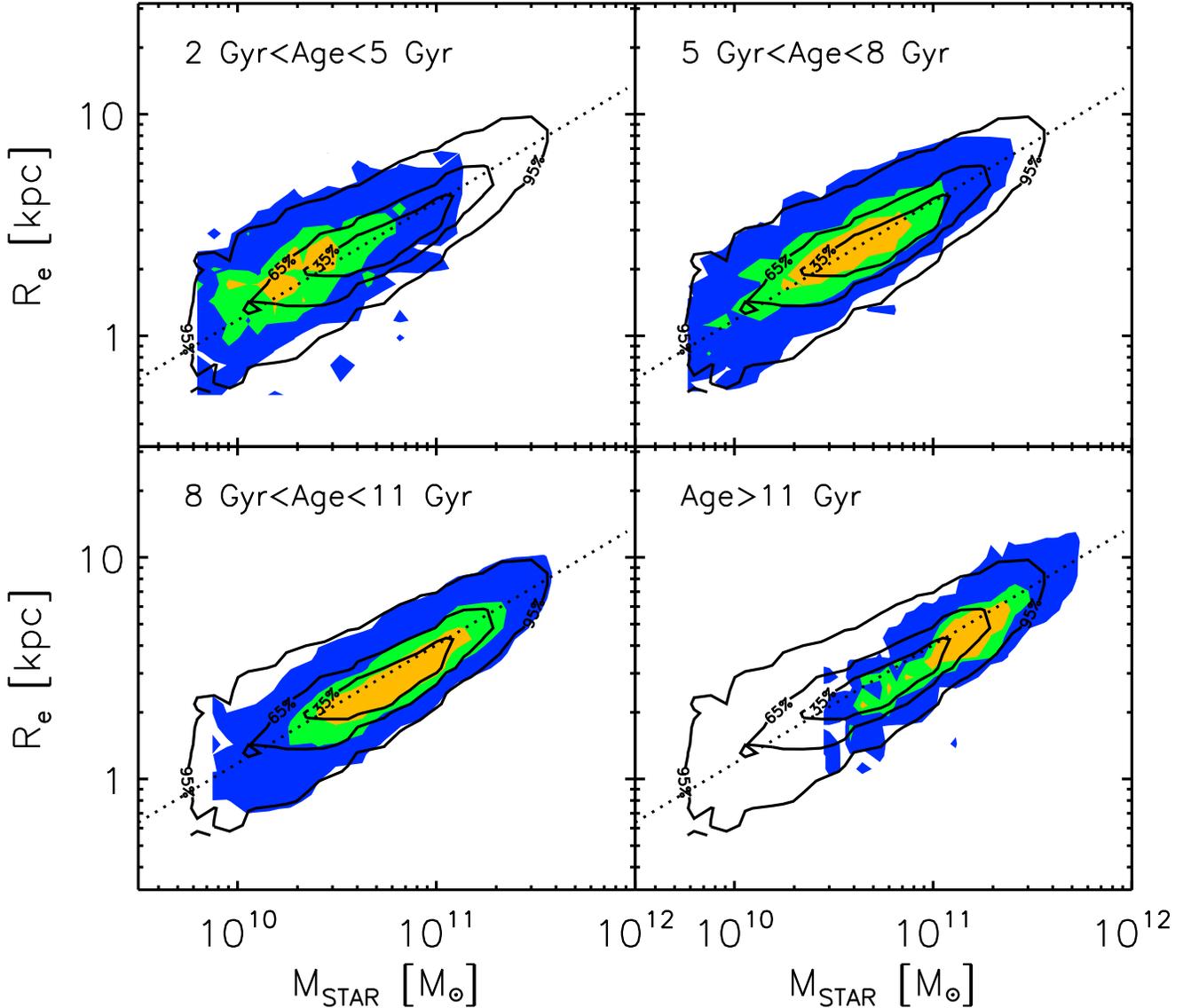}
 \caption{Size-stellar mass relation for the galaxies
 in our sample with ages as labeled in the upper left corner of each
 panel. The solid lines bracket the
 regions containing 30\%, 65\%, and 95\% of the full sample, respectively.
 The orange, green, and blue coloured regions show the
 corresponding distributions when we restrict the sample to narrow bins in (luminosity-weighted)
 age. Older galaxies shift to higher \re, and \mstar, but are not
 offset from the relation defined by the full sample, whereas the
 youngest objects tend to be offset towards larger \re\ and smaller \sis.
 \label{fig|RevsMstarGlobal}}
\end{figure*}

\begin{figure*}%[ht!]
\includegraphics[width=17.5truecm]{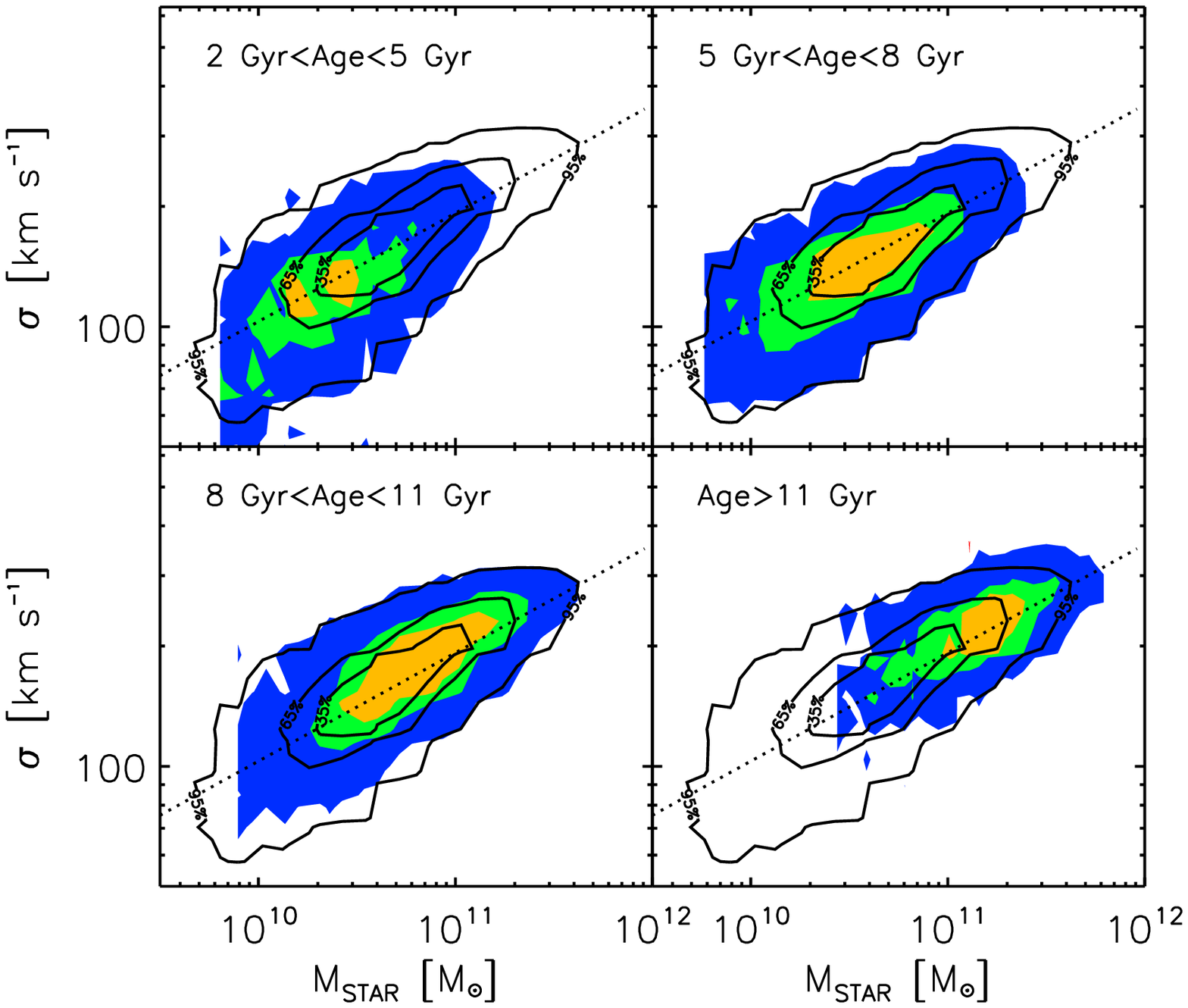}
 \caption{Corresponding velocity dispersion-stellar mass
 relations with same format as Figure~\ref{fig|RevsMstarGlobal}.
 Similarly, older galaxies shift to higher \sis and \mstar, but are not
 offset from the relation defined by the full sample, whereas the
 youngest objects tend to be offset towards smaller \sis.
 \label{fig|SigmavsMstarGlobal}}
\end{figure*}

\citet{ShankarBernardi} used a sample of about 48,000 early-type
galaxies from the Sloan Digital Sky Survey (SDSS; \citealt{York00})
to show that older galaxies have smaller half-light radii \re\ and
larger velocity dispersions \sis\ than younger ones of the same
stellar mass \mstar. Specifically, they found that, using the
age-corrected luminosity \Lcorr\ as a proxy for \mstar,
galaxies with age $\sim 11$~Gyrs below \Lcorr$\sim 10^{11}\, $\lsun\
have \re\ smaller by 40\% and \sis\ larger by 25\%, compared to
galaxies that are 4~Gyrs younger. The sizes and velocity dispersions
of more luminous galaxies vary by less than $15\%$, whatever their
age, implying a significant \emph{break} in the \re--\Lcorr\ and
\sis--\Lcorr\ relations at high \Lcorr. The galaxies in their sample
have been carefully selected to be early-type galaxies, and they are
all characterized by much larger sizes and lower densities than
their higher redshift massive counterparts. Also, the differences
between the sizes of old and young galaxies reported by
\citet{ShankarBernardi}, are far less than expected from a simple
monolithic model evolution with \re$\propto (1+z)^{-1}$ at
fixed stellar mass, which would result if the galaxy density is
proportional to the density of the universe.

Our goal in what follows is to compare these findings data with
hierarchical models of galaxy formation and evolution to understand
how well they can reproduce the data.
In Section~\ref{sec|data} we first revisit the
main observational results regarding the size-mass
relation in local ellipticals, showing that the very massive galaxies
of any age follow similar size-stellar mass relations in shape,
normalization, and dispersion around the mean.
Section~\ref{sec|models} compares our measurements with the publicly
available SAM of \citet[][B06 hereafter]{Bower06}, which has
successfully reproduced several statistical properties of galaxies.
The B06 model, which
is based on the Millennium Simulation of the dark matter distribution
\citep{Springel05}, provides the sizes of spheroids at any epoch,
thus enabling a direct comparison between the predicted and observed
size-mass evolution.
We use the SAMs to discuss how the number and type of mergers
(dry or wet) scale with final stellar mass.
In particular, we show that dry mergers, defined to be mergers
between gas-poor progenitors, might be good candidates for erasing
the effects of a monolithic collapse and producing a rather flat
size-age relation, similar to what observed in the data.
Our conclusions are in \S~\ref{sec|conclu}, where we also discuss
some serious failures of the models.

Throughout this paper we adopt the cosmological parameters
$\Omega_m=0.30$, $\Omega_\Lambda=0.70$, and $h\equiv H_0/100\, {\rm
km\, s^{-1}\, Mpc^{-1}}=0.7$.

\section{DATA AND MEASUREMENTS}
\label{sec|data}

\subsection{The sample}
We use the SDSS-based sample of early-type galaxies from \cite{Hyde09a}.
This sample was constructed in a way to maximize
the contribution of elliptical galaxies as briefly
described here (see the recent work
by \citealt{Bernardi09b} for further details). The galaxies
in this sample are very well described by a deVaucouler profile
in both the $g$ and $r$ bands (\emph{fracDev}=1), have ''early-type''
spectrum, with eClass$<0$ (to minimize later-type contamination), and have an
additional cut in axis ratio of $b/a>0.6$. These galaxies were
also selected to have velocity dispersion \sis\ higher than 60 \kms, close to the dispersion limit of the SDSS spectrograph, and less than 400 \kms, to
avoid contamination from double/multiple superpositions (Bernardi et al. 2006, 2008).
%Also
%note that the contamination from low star-forming galaxies
%should be small in this sample. The SDSS-DR6 in fact reports velocity dispersions only when the signal-to-noise ratios are greater than 10 and the status flag is equal to 4, a classification
%which tends to exclude galaxies with emission lines.
The resulting sample, which contains about 47,300 early-type galaxies, is
distributed within the redshift range $0.013 < z < 0.3$, which
corresponds to a maximum lookback time of 3.5 Gyr.
Based on the recent, detailed
analysis performed by \citet{Bernardi09b}, this
sample has minimal contamination by disky S0 galaxies,
which makes it an ideal catalog to compare with
galaxy evolution models of massive spheroids.
The galaxies in the sample have apparent magnitudes
$14.5\lesssim m_r \lesssim 17.5$ (based on deVaucouleur fits to
the surface brightness profiles).  (The SDSS photometric parameters for
these objects have been corrected for known sky subtraction problems
which affect bright objects.)  Estimated stellar masses and ages for
these objects are from \citet{Gallazzi05}.
These are based on running a likelihood analysis of the spectra
that returns a mass-to-light ratio \mstar$/L_z$ (defined for a \citealt{Chabrier03} initial
stellar mass function), which is in turn converted
to a stellar mass using the SDSS petrosian $z$-band restframe magnitude
(see \citealt{Bernardi09b} for a comprehensive discussion
and comparison of such stellar mass estimates with other
methods).
Our main results do not change if we use the ages
published by \citet{Bernardi06}, which were computed by fitting the
\citet{Thomas05} $\alpha$-enhanced models to the Lick index absorption
features.  The age estimates of \citet{Jimenez07}, derived from single
stellar population spectral fitting, using the MOPED algorithm
(\citealt{Heavens00}), also yield similar results. Also,
a possible bias might be introduced by the youngest galaxies
in the sample with ages $<5 Gyr$ (the young age could be due to weak star formation
which makes the galaxy to appear relatively young), comparable to the
lookback time of the sample and close to the average error in age estimates.
However, we have checked that cutting out these galaxies and
repeating the analysis does not minimally alter the overall conclusions
of the paper.

Finally, we note that the age and stellar mass estimates come from
the same algorithm, so they have correlated errors.  However, this
does not bias the results which follow
\citep[for details, see][]{Bernardi09, ShankarBernardi}.
%In the following, we will consider only galaxies with ages above 5 Gyr.

In what follows we wish to study the bulk of the early-type
population.  However, e.g., \citet{Bernardi09} has argued that
Brightest Cluster Galaxies (BCGs) had unusual formation histories,
so, following \citet{ShankarBernardi}, we remove them from our sample.
Our final sample is composed of $\sim 45,700$ galaxies.

\begin{figure*}%[ht!]
\includegraphics[width=0.45\hsize]{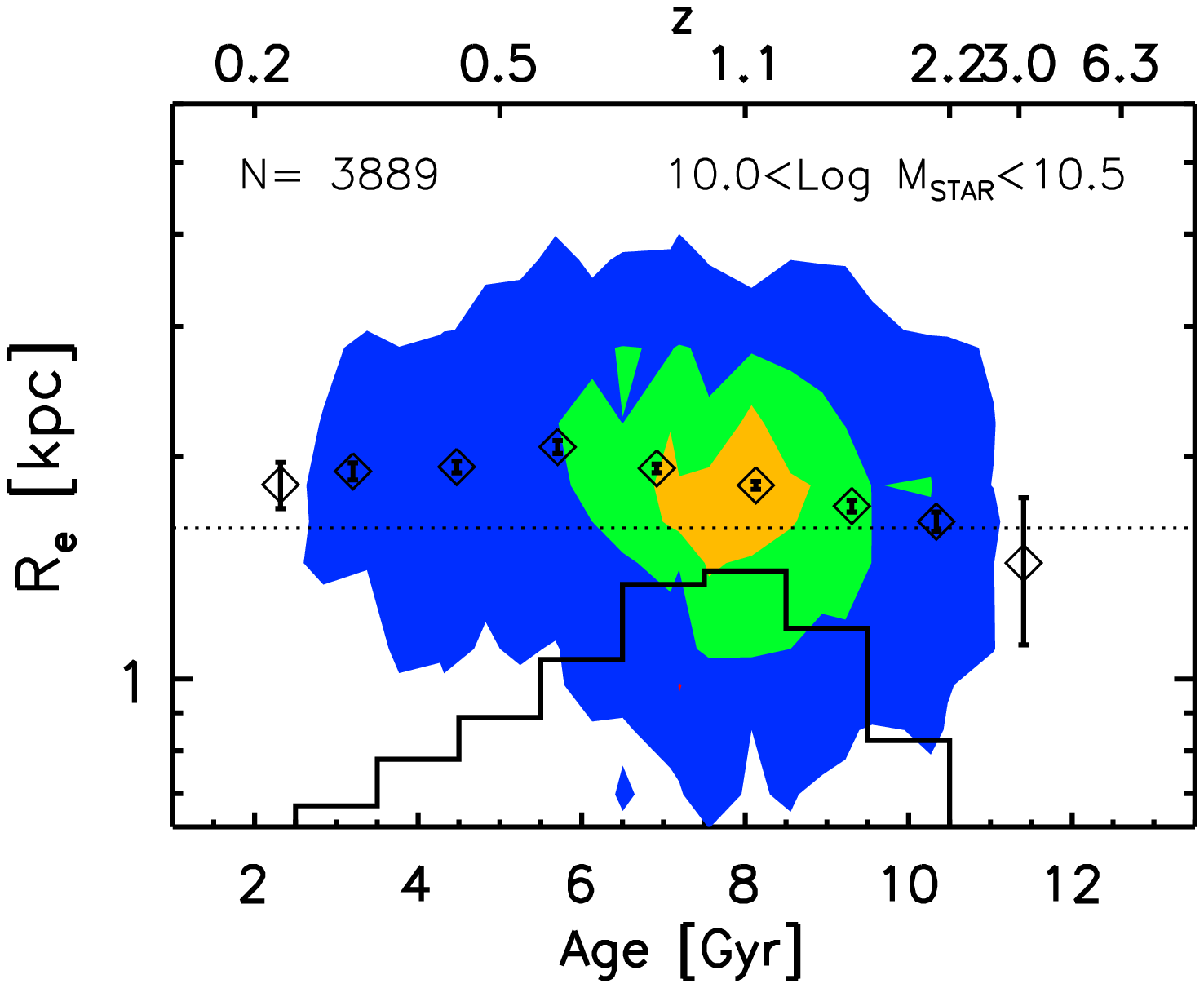}
\includegraphics[width=0.45\hsize]{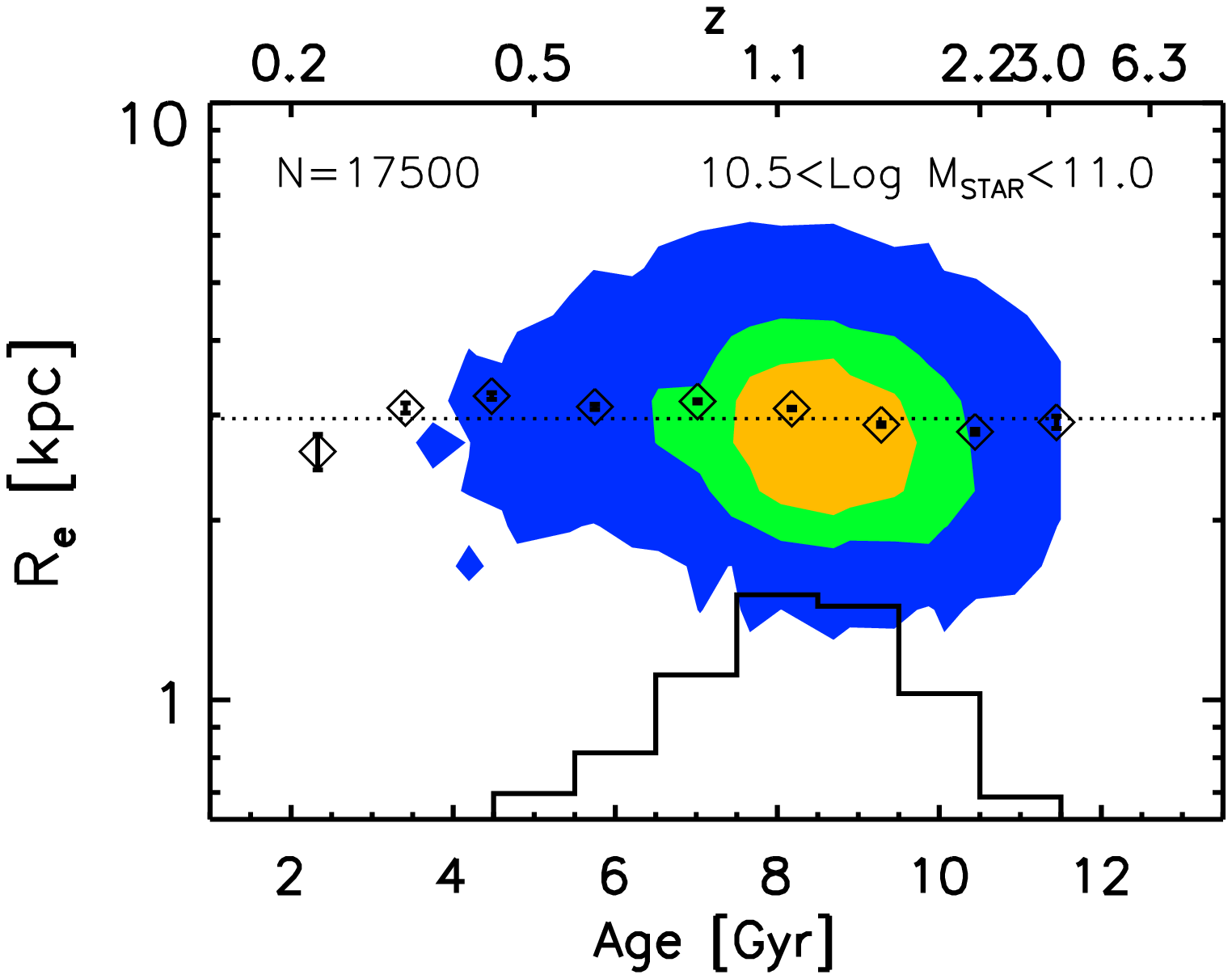}
\includegraphics[width=0.45\hsize]{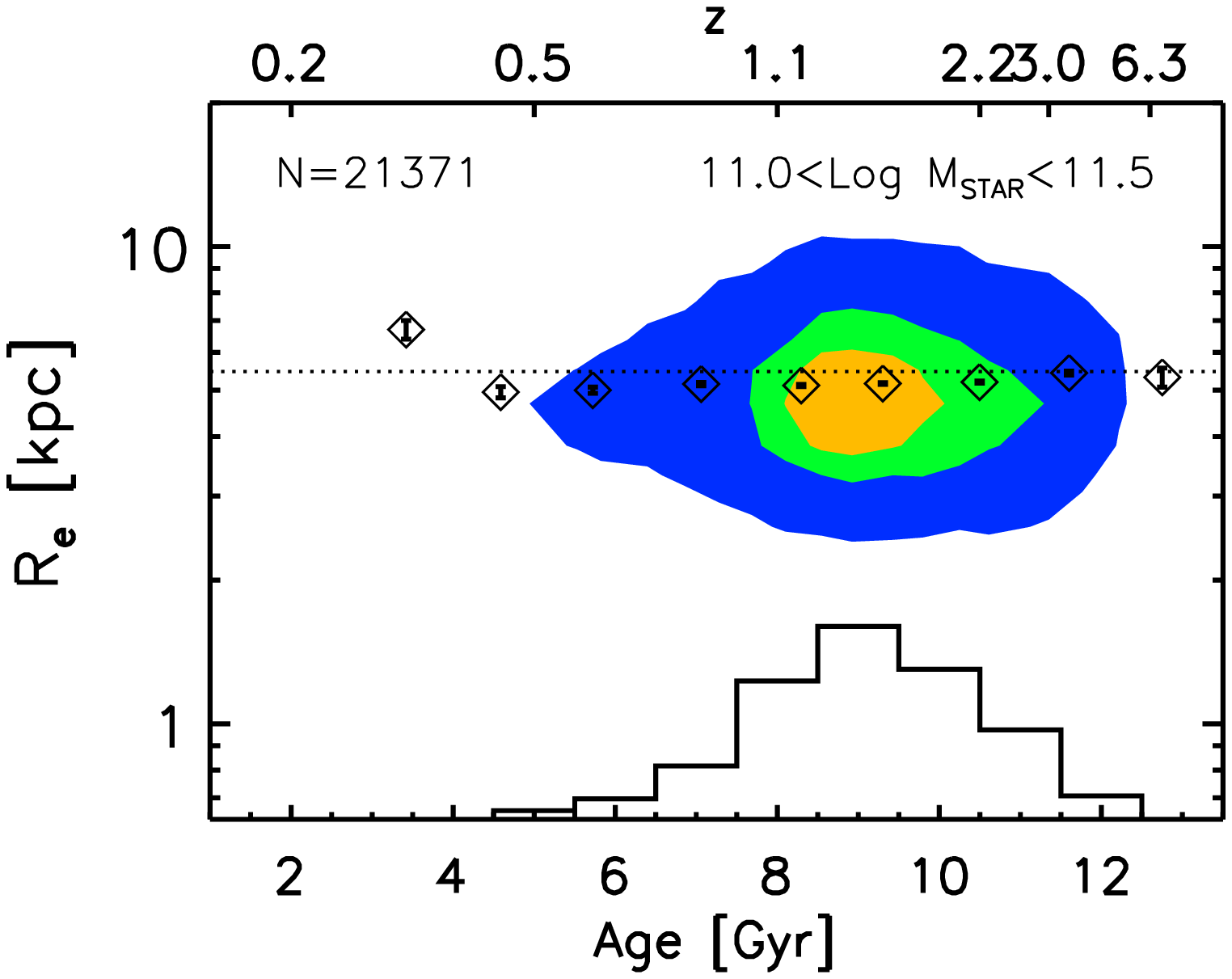}
\includegraphics[width=0.45\hsize]{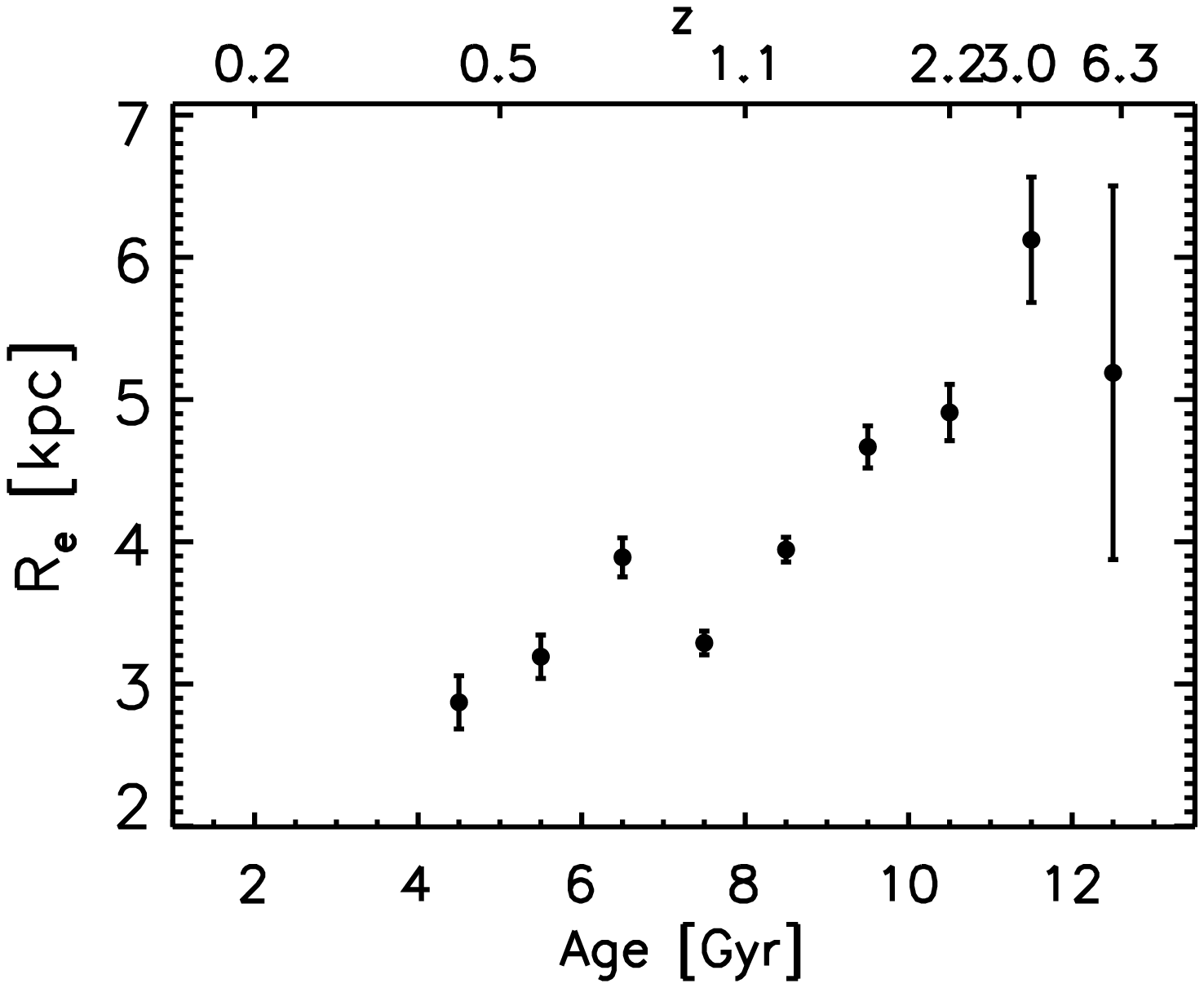}
{\caption{Mean effective radius
as a function of age for a given subsamples defined by stellar mass;
histograms show the age distribution of each subsample.
The bottom right panel shows the mean effective
radii found by averaging over 100 simulations
in which each time a subsample of 1000 galaxies is randomly drawn from
the catalogue: a mere random selection of galaxies from the parent
sample does not reproduce the generally flat $R_e-$age relation at
fixed stellar mass. The dotted line in each panel
shows the mean size for
the stellar mass bin considered,
as predicted by the global \re-\mstar\ relation
(dotted line in Figure~\ref{fig|RevsMstarGlobal}).}
\label{fig|ReFixedMstar}}
\end{figure*}

\begin{figure*}%[ht!]
\includegraphics[width=0.43\hsize]{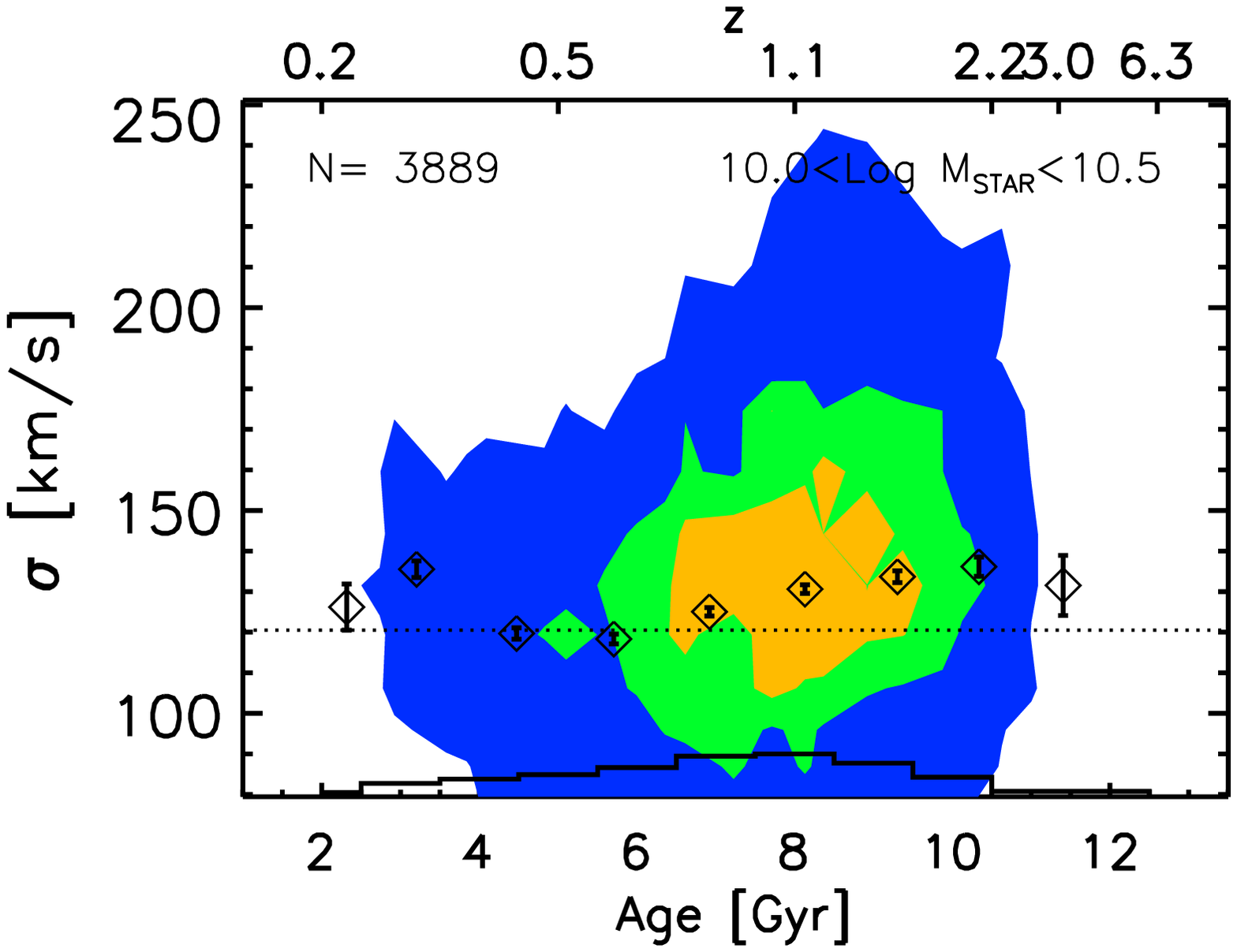}
\includegraphics[width=0.43\hsize]{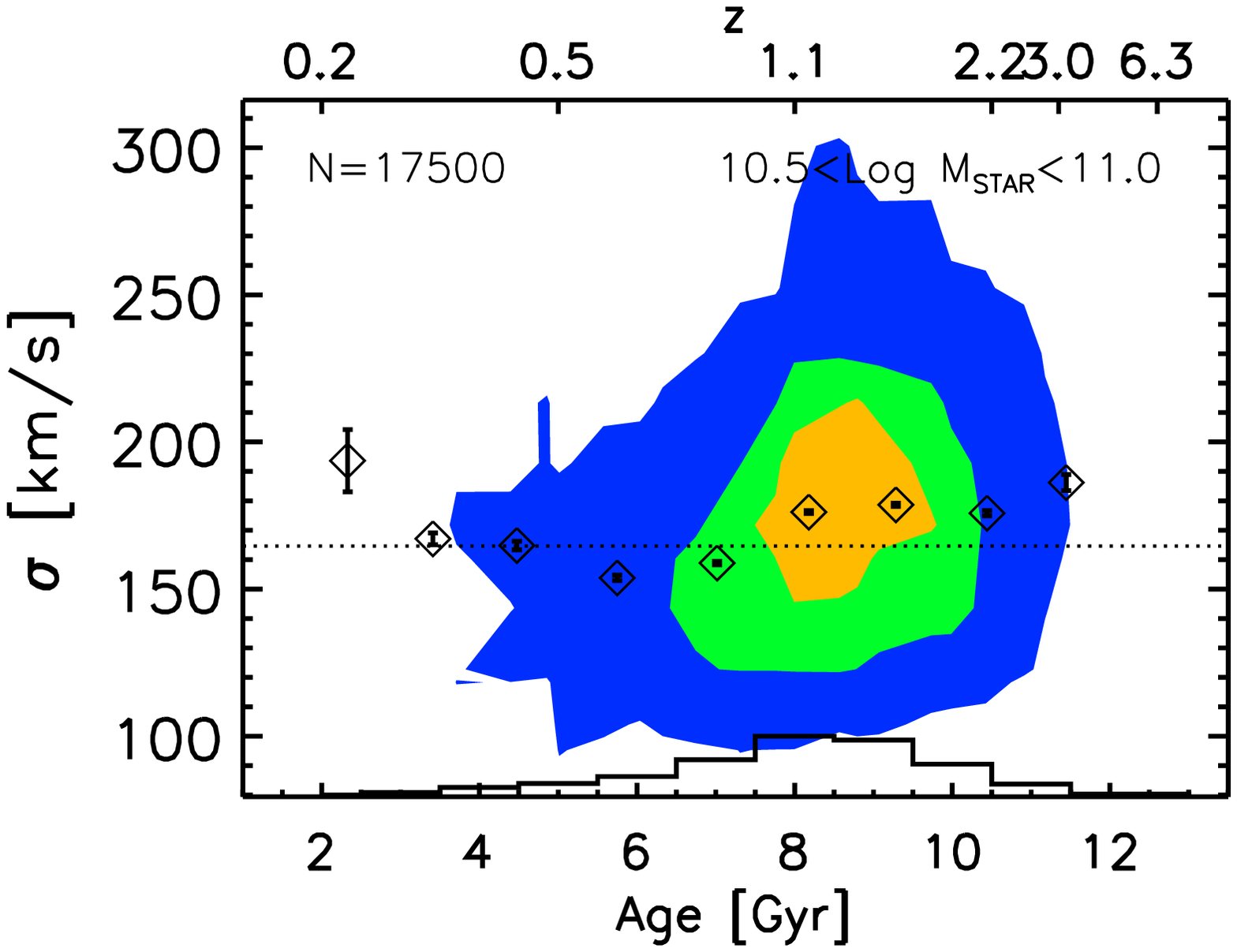}
\includegraphics[width=0.43\hsize]{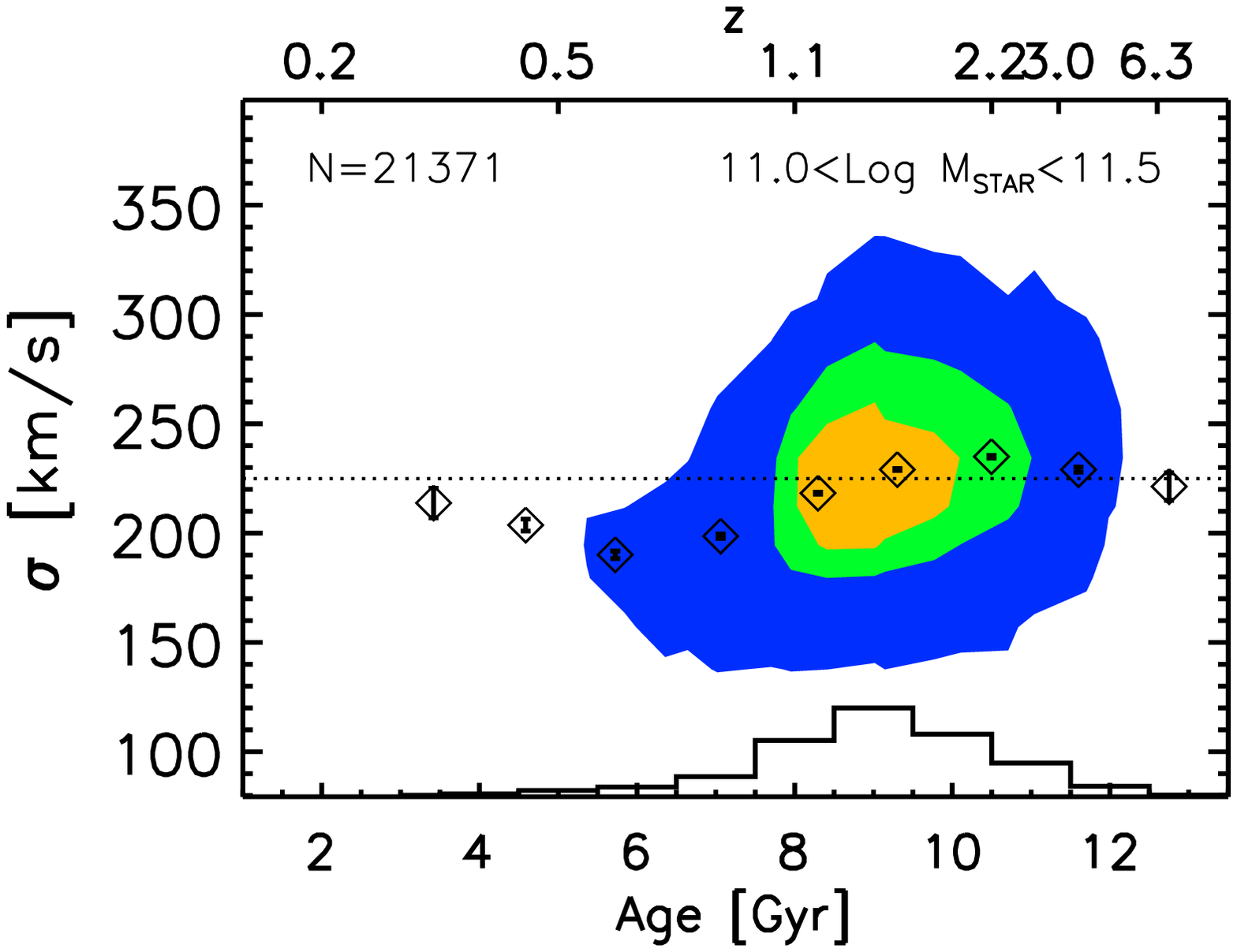}
\includegraphics[width=0.43\hsize]{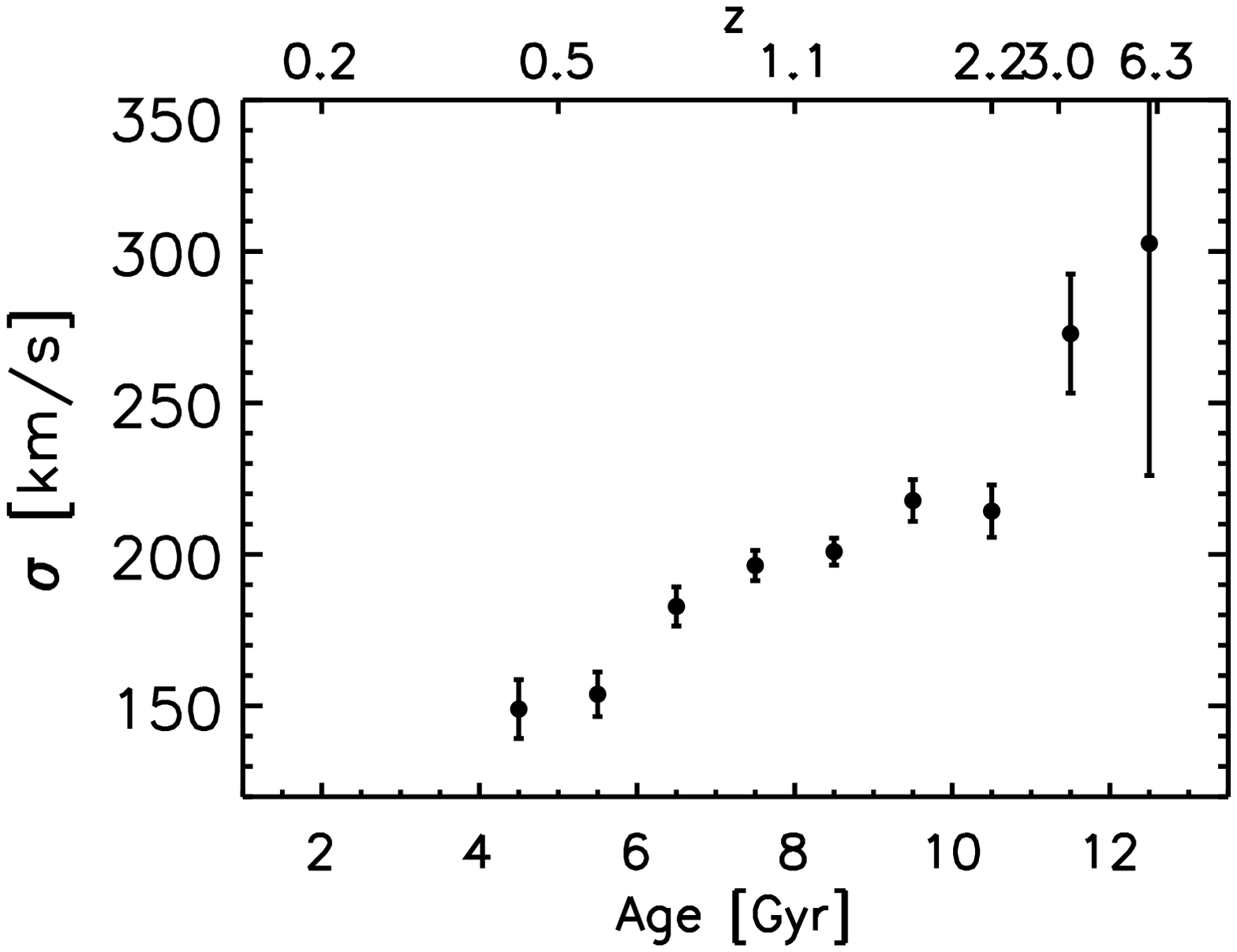}
{\caption{Same pattern as Figure~\ref{fig|ReFixedMstar}. Mean velocity dispersion
as a function of age for a given subsamples defined by stellar mass, with
histograms showing the age distribution of each subsample.
The bottom right panel shows the mean velocity dispersions
from random selection of galaxies from the parent
sample. The dotted line in each panel
shows the mean velocity dispersion for
the stellar mass bin considered,
as predicted by the global \sis-\mstar\ relation
(dotted line in Figure~\ref{fig|SigmavsMstarGlobal}).}
\label{fig|SigmaFixedMstar}}
\end{figure*}

\subsection{Results}\label{sec|results}

It is well known that more massive early-type galaxies have larger
half-light radii.  The dotted line (same in all four panels) in
Figure~\ref{fig|RevsMstarGlobal} shows this relation in our
dataset, obtained from a linear fitting  $\log (R_e/5 kpc)=(-0.63\pm 0.01)+(0.53\pm 0.01)\log (M_{\rm STAR}/10^{10}\, M_{\odot})$,
in very good agreement with previous results from \citet{Shen03}.
Notice also that the spread in size at fixed mass is approximately
$\sim 0.6$ dex for masses below \mstar$\sim 3\times 10^{10}$\msun, decreasing at large mass.
This behavior in the scatter as a function
of stellar mass was already noted by \citet{Shen03},
and we here further suggest, after
\citet{ShankarBernardi} and the discussion below, that this
is mainly induced by the different gradients in the size-age relation in different stellar mass bins.
The solid contours, the same in each panel, bracket the regions
containing 30\%, 65\%, and 95\% of the full sample, respectively.

The orange,
green, and blue coloured regions show the
corresponding distributions when we restrict the sample to narrow bins in (luminosity-weighted)
age.  Notice that the older galaxies tend to populate the high-mass
end of this relation (the shaded regions peak at higher \mstar\ in
the bottom right panel than in the top left), in qualitative agreement
with the notion of downsizing \citep[e.g.,][]{Cowie96,Heavens04}.
To proceed in a more accurate
analysis, we have refitted the size-mass relation for the subsamples
of galaxies considered in Figure~\ref{fig|RevsMstarGlobal}. We found
that while younger galaxies follow a significantly
shallower correlation, $R_e\propto$ \mstar$^{0.48\pm 0.01}$,
the oldest ones follow
a steeper relation, with $R_e\propto$ \mstar$^{0.65\pm 0.01}$.
These behaviors are mainly caused
by the fact that below \mstar$\lesssim 10^{11}\, $\msun, older galaxies tend to
gradually have
sizes that are up to a factor of about two smaller than younger ones
of the same \mstar.  Above $10^{11}\, $\msun, the sizes are instead similar,
whatever the age.

Figure~\ref{fig|SigmavsMstarGlobal} shows a
similar analysis of the velocity dispersions \sis\ rather than the
sizes.  As for the sizes, the overall scatter in \sis\ at fixed
\mstar\ is independent of age, but decreases with increasing \mstar, lending
further support to our above suggestion that
what is actually driving the mass-dependent
scatter in the size-mass relation is correlated to differences
in ages (see also \citealt{ShankarBernardi}).
And, to lowest order, the mean velocity dispersion-stellar mass
relation, which we fit as $\log (\sigma/200\, km/s)=(-0.288\pm 0.014)+(0.27\pm 0.01)\log(M_{\rm STAR}/10^{10}\, M_{\odot})$,
is the same in all age bins -- the
primary trend being that older galaxies are shifted to larger \mstar\
values.  A more detailed analysis yields that, at the low mass end, young
galaxies (which had slightly larger sizes) have slightly lower velocity
dispersions, thus causing a slight steepening of the relation.

Figures~\ref{fig|ReFixedMstar} and ~\ref{fig|SigmaFixedMstar} show another view of the correlations
between size, mass and age: the size-age relation for bins in \mstar.
In agreement with the previous Figures and what discussed in \S~\ref{sec|intro} (see also \citealt{ShankarBernardi}), the lower mass galaxies
(upper left panels)
have sizes and velocity dispersions which show some significant trends with age, with older galaxies having smaller \re\ and larger \sis. At the top
of each panel we also indicate, for better reference to the
models discussed below, the redshift corresponding
to the lookback time equal to the age.
The dotted lines in the Figures show the mean values of size or
velocity dispersion predicted by the global \re-\mstar\ and \sis-\mstar\ relations
(dotted lines in Figures~\ref{fig|RevsMstarGlobal} and \ref{fig|SigmaFixedMstar})
for the stellar mass defined in each panel.
It is clear that while the lowest
mass bins present a significant
gradient with age (opposite sign for sizes and velocity dispersion),
this trend progressively disappears when moving to more massive
galaxies, with almost no correlation
between size and age for the most massive ones.

Nevertheless, more massive objects are offset to larger sizes.
As a result, when averaged over all masses, older objects
have larger sizes, except possibly for the oldest galaxies.
The bottom right
panels of Figures~\ref{fig|ReFixedMstar} and
\ref{fig|SigmaFixedMstar}
show the results of this exercise. When
randomly selecting galaxies from the whole sample
(the points are averages
over 100 realizations with 1000 points
each), we find an increasing
size and velocity dispersion with increasing
stellar mass.
Thus, the size-age relation is almost entirely due to the
size-mass and age-mass correlations (massive objects are older and
larger).  This is analogous to the color-magnitude relation being
entirely due to the correlations between color and luminosity with
velocity dispersion \citep{Bernardi05}.

Similarly, at fixed \mstar, there is little correlation between
\sis\ and age, but, because more massive galaxies are offset to
larger \sis, the result of averaging the \sis\ -age relation over all
\mstar\ yields a strong trend:  the oldest galaxies have the highest
values of \sis\ \cite[e.g.,][]{Bernardi05}.

\subsection{Some implications}

In a strict monolithic scenario for galaxy formation in which the
age of the stars reliably traces the time the galaxy was assembled,
older galaxies should have smaller sizes than younger galaxies of
the same stellar mass.  This is because objects which assembled
earlier did so in a denser universe.  If the galaxy density is
proportional to the density of the universe at the time of assembly,
one expects \re\ $\propto (1+z_{\rm form})^{-1}$ at fixed stellar mass.
(One might expect the actual scaling to be even stronger,
since dissipation in the baryonic clumps from which the stars form
is expected to be more efficient when the density is higher, so
objects which form at higher redshift should be even more compact
and have even higher velocity dispersions.)
However, the sizes of old and young galaxies in our sample are
much weaker functions of age.
This rules out models in which star formation and assembly are
concurrent, and galaxies passively evolve thereafter.

On the other hand, as reviewed in Section~\ref{sec|intro}, a large portion of
high redshift galaxies {\em are} observed to be more compact than their counterparts
at low $z$.
Therefore, some process must have altered the sizes and stellar masses
of galaxies since they formed.  Our results suggest that, whatever
the mechanism, it must be ``fine-tuned'' so as to yield the
essentially flat size-age and $\sigma$-age relations (at fixed \mstar)
shown in Figures~\ref{fig|ReFixedMstar} and \ref{fig|SigmaFixedMstar}.

\begin{figure*}%[ht!]
\includegraphics[width=0.9\hsize]{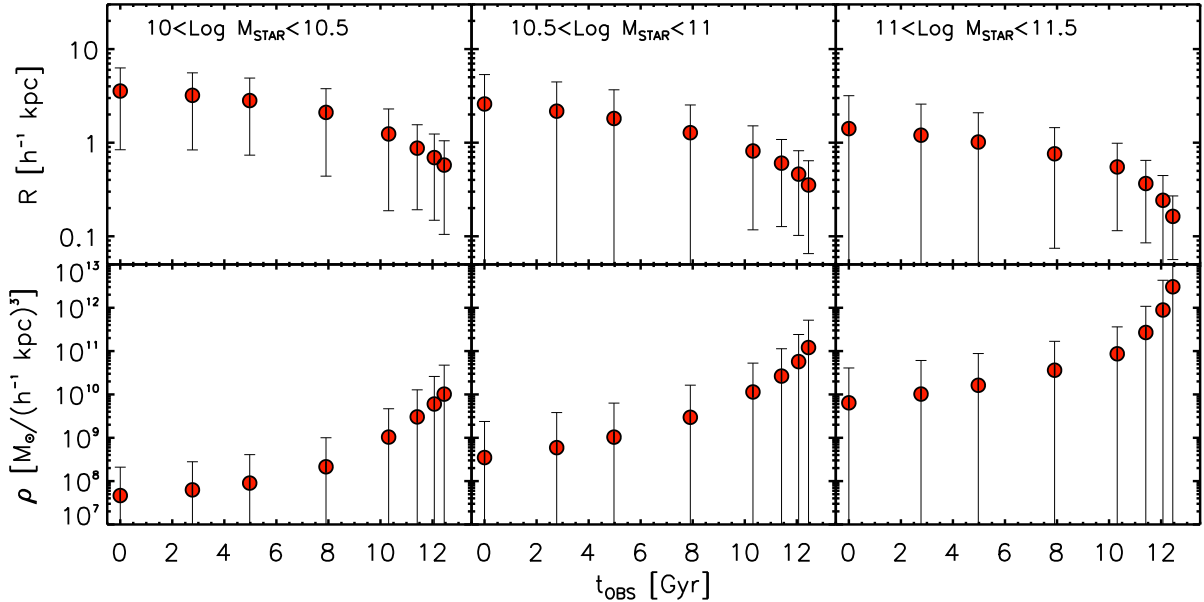}
{\caption{Predicted sizes (\emph{upper} panels) and densities
(\emph{lower} panels), as well as the dispersions around the mean
for galaxies of the same stellar mass, as labeled in each column,
but identified at different epochs, in the Bower et al. (2006) model.
On average, galaxies of the same stellar mass are predicted to be
significantly smaller and denser at early times.}
\label{fig|RezModelsFixedMstar}}
\end{figure*}

\section{COMPARISON WITH GALAXY FORMATION MODELS}
\label{sec|models}

In this Section we compare our observational data with the
predictions of semi-analytic galaxy formation models.
See \cite{Parry08} and \PapI\ for a detailed description and
comparison of such models, which follow the cosmological co-evolution
of dark matter halos, subhalos, galaxies and supermassive black holes
within the concordance $\Lambda$CDM cosmology.
Briefly, these models track disk and bulge components of each
object as it evolves.  B06 also output the half mass radius of the
spheroid component, so this is the model we use as a reference in what
follows.

We will be interested in the masses, (half-mass) sizes, and formation
histories of the early-types galaxies in these models.
We classify a galaxy as early-type if it has
 $M_{\rm bulge}-M_{\rm total}<0.4$,
where $M_{\rm bulge}$ and $M_{\rm total}$ are the predicted $B$-band
magnitude of the bulge and of the whole galaxy, respectively. This
cut selects galaxies with a bulge-to-total ratio $B/T\gtrsim 0.7$, which
is the minimum $B/T$ characterizing the galaxies in the
\citet{Hyde09a} sample. Note that
this cut in $B/T$ preferentially selects
spheroid-dominated galaxies, while a more common cut of $B/T>0.5$,
comparable to a cut in concentration of $C_r>2.86$ (e.g., \citealt{GonzalezSAM08}),
allows for a large contamination of disky galaxies (see the recent
analysis by \citealt{Bernardi09}). The latter type of galaxies
might have had quite different formation histories than the ones
of interest to this paper. In fact, a large fraction
of S0 galaxies in the B06 model grow their
bulge via disk instability. Also, as further discussed
in Shankar et al. (2009c), the mock galaxy sample considered here
covers a wide variety of stellar masses
and luminosities, as in the observed one. Nevertheless,
as discussed by \citet{GonzalezSAM08}, \PapI\ and further below, the
model has a tendency to produce broader distributions
at fixed size and/or stellar mass than those actually
observed. We believe that
this effect is mainly due to somewhat inappropriate physical
recipes more than inadequate selection cuts in the model (see the detailed analysis by \citealt{GonzalezSAM08} on this exact point).

We identify the formation epoch of an early-type galaxy as the first
time along the merger tree that the progenitor becomes an early-type.
Note that this classification of early-types and their formation
does not make any assumptions about the color or the star formation rate.
In the following, we will present results based on the full sample of
early-type galaxies, regardless of whether they end up being central
or satellite galaxies at $z=0$.  However, because we have removed BCGs
from the data, we have checked that our basic result, of a flat size-age
relation at fixed stellar mass, is still conserved if we remove the
objects which are centrals at $z=0$ in the model halos.

\subsection{Higher densities at early times?}
Before we present a more direct comparison of these models and
the data shown in the previous figure, it is interesting to
see if the models are consistent with the notion that objects
at high redshift are denser, for the reasons discussed in the
Introduction.  Figure~\ref{fig|RezModelsFixedMstar} shows the
sizes (top) and densities (bottom) as a function of lookback
time, for objects of fixed stellar mass \mstar\ at each lookback time.
The three panels show these predictions for three choices of \mstar.
Notice that high-redshift objects are smaller (and hence denser) than
their counterparts of the same \mstar\ at later times -- in qualitative
agreement with observations and expectations.

However, notice that the models predict the typical size to
{\em decrease} as \mstar\ increases.
This is grossly discrepant with observations, and suggests that there
is considerable room for improvement with regards to how the models
assign sizes -- this was also noted by \citet{GonzalezSAM08} and \PapI.
We further discuss in \PapI\ the actual successes and failures
of the present model by making use of a combined comparison
with the size and mass distributions as observed
in SDSS.
In this paper, however, we are less interested in the form of the correlation
between size and mass --- we are more interested in checking if a
``chaotic'' formation model, such as a hierarchical one, in which
early-type galaxies form and grow continuously, can reproduce the
fact that the size-mass relation is independent of age -- as observed
for today's massive galaxies, irrespective
of whether the final size is actually the one observed
for that bin of stellar mass.

\begin{figure*}%[t]
\includegraphics[width=0.9\hsize]{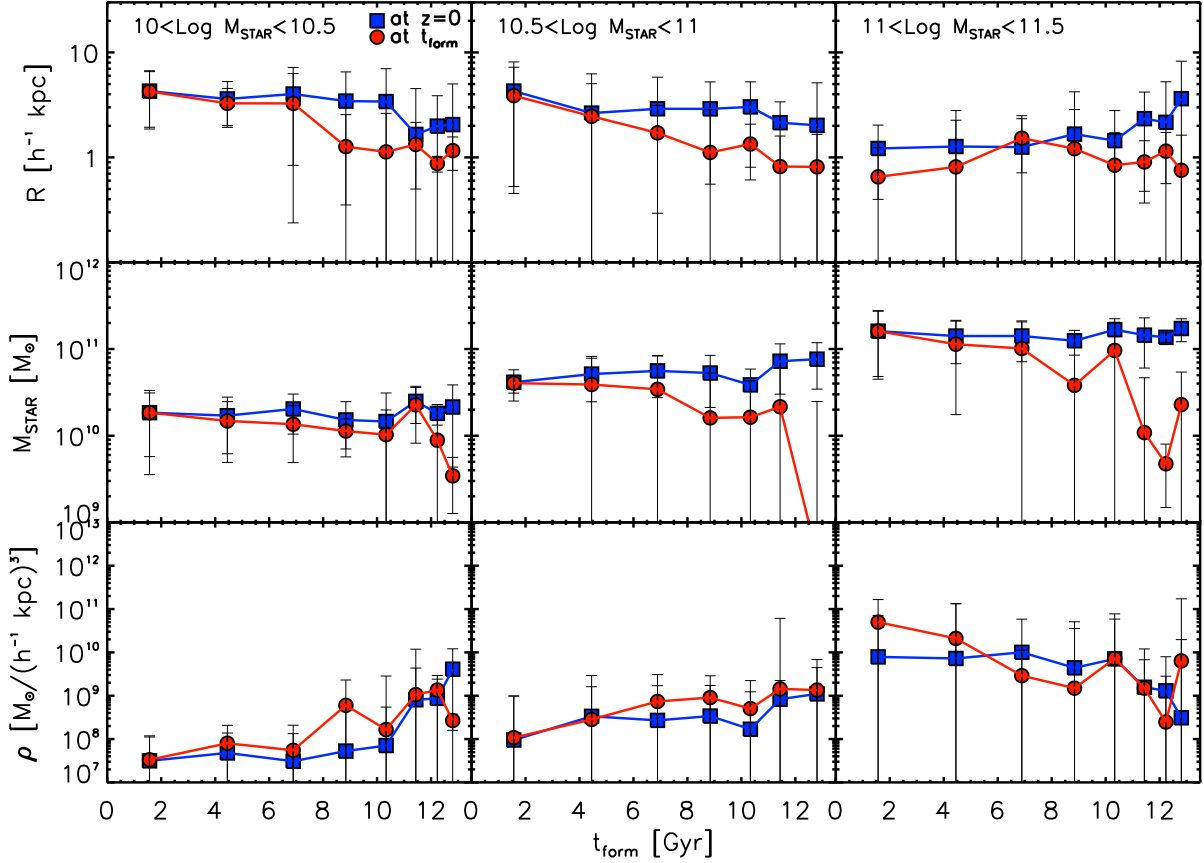}
{\caption{Mean galaxy parameters (spheroid size, stellar mass and
stellar density) as a function of age, for objects in narrow
mass bins as labeled.  Each panel shows an average over one
hundred merger histories in the B06 model. The \emph{blue squares} refer to the properties
of galaxies (size, mass, or density) at $z=0$, while
the \emph{red circles} refer to the properties the same
set of galaxies had at their formation epoch, $t_{\rm form}$.}\label{fig|RezHistory}}
\end{figure*}

\begin{figure*}%[ht!]
\includegraphics[width=0.9\hsize]{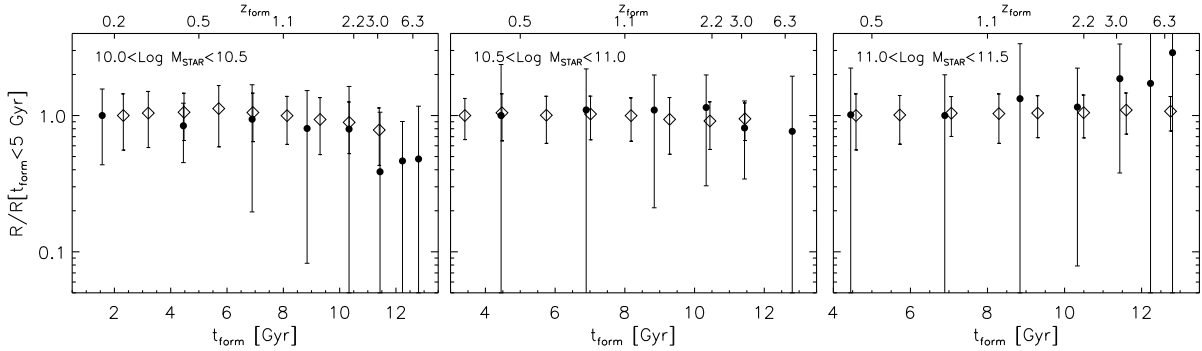}
{\caption{Present day sizes as a function of formation time,
          normalized by the size when the formation time was
          very recent in the SDSS data (\emph{open diamonds},
          rescaled from Figure~\ref{fig|ReFixedMstar})
          and the SAMs (\emph{solid circles}, rescaled version
          of the blue circles of Figure~\ref{fig|RezHistory}).
\label{fig|compactComparison}}}
\end{figure*}

\subsection{Mass dependent evolution of sizes and masses}

Figure~\ref{fig|RezHistory} shows the model predictions in a
format that is closer to that shown in the previous section.
We here plot the median sizes, stellar masses and densities
(along with their dispersions), averaged over 100 merger histories\footnote{We have checked that the main results of this paper do not depend significantly on the specific
number of random merger histories adopted. Increasing this number by a factor
of a few yields essentially equal results both in the mean trends
and broadness of distributions around the mean.},
of galaxies with stellar mass at $z=0$ in the mass bin indicated at
the top of each column. In each case the trees are followed back in
time, choosing the most massive early-type progenitor, until this is
no longer possible. The blue squares refer to the properties
of galaxies (size, mass, or density) at $z=0$, while
the red circles refer to the properties the same
set of galaxies had at their formation epoch, $t_{\rm form}$
(defined as lookback time in Gyr).
The blue squares in the top panels, which show the $z=0$ size-age
relation for galaxies in narrow bins in \mstar, can be directly compared
with the measurements shown in Figure~\ref{fig|ReFixedMstar}, assuming
the age of a galaxy is a good proxy for its formation
epoch $t_{\rm form}$.
Notice that, at fixed \mstar, the size-age relation is weak,
with a slight tendency for old objects to be smaller in the
smallest \mstar\ bin, but to be larger in the largest \mstar\ bin.
Indeed, the models suggest that, at small \mstar, older objects have
slightly smaller sizes whereas the opposite is true at higher
\mstar.  The sense of these weak trends at fixed \mstar\ is also
in qualitative agreement with our measurements in the SDSS, and
may be considered a significant success of the models.
However, we stress that the models predict {\em smaller} sizes at
large \mstar, whereas the data shows the opposite trend.  As a
result, the overall $R_e-$\mstar\ relation in these models is
grossly discrepant with that in the SDSS.  (This is essentially the
same problem we found in Figure~\ref{fig|RezModelsFixedMstar};
the only difference is that there the galaxies in a given panel
were selected to have the same \mstar\ at all lookback times, whereas
here they have the same \mstar\ only at $z=0$, and we then study their
progenitors at earlier times.)

\begin{figure*}%[ht!]
\includegraphics[width=0.9\hsize]{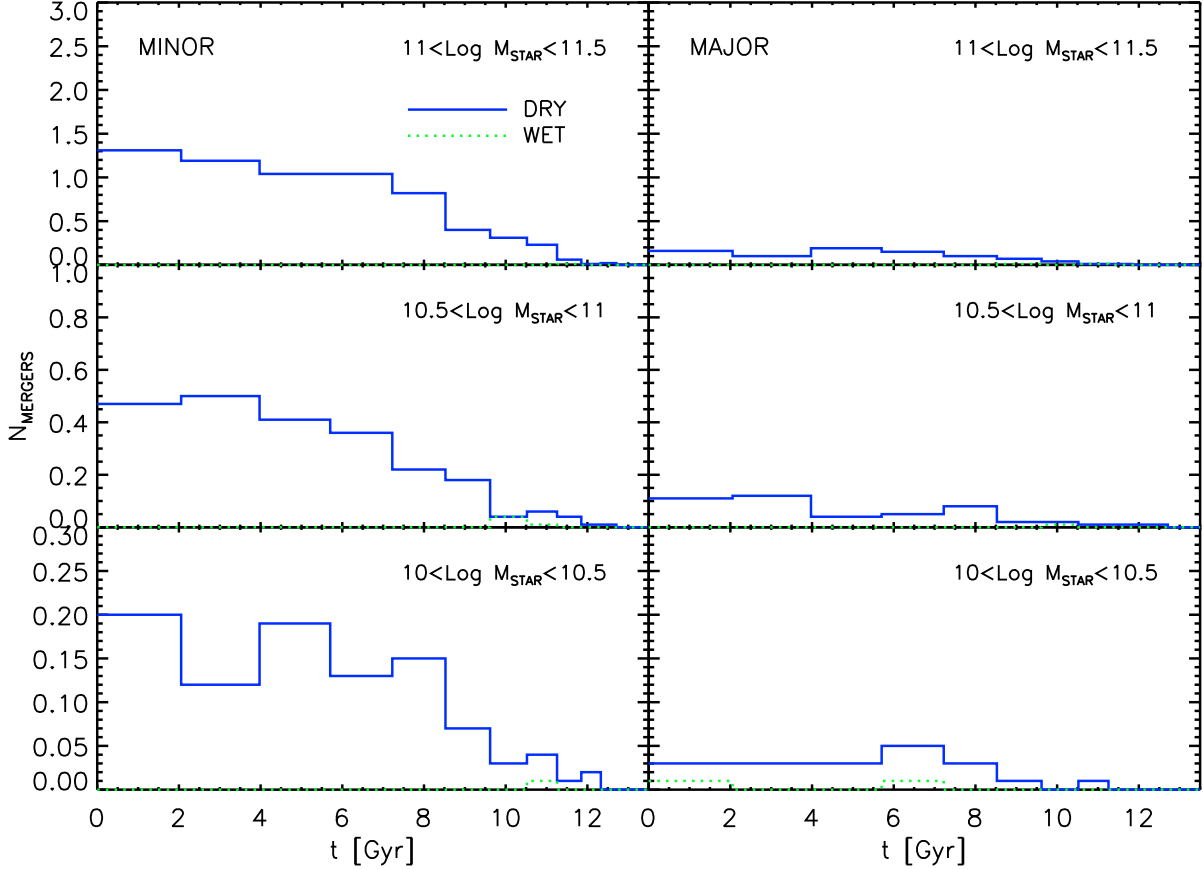}
{\caption{Comparison between the mean number of wet and dry mergers
per Gyr, averaged over 100 galaxies,
extracted from the merger trees of the Bower et al. (2006;
\emph{blue} lines)
catalogues. Each row shows the merger history, averaged
over 100 realizations, of galaxies with stellar mass at $z=0$ in
three different mass bins, as labeled. In the left column we plot
the mean number of minor mergers, with mass ratio $<1:3$, while the
right column shows the mean number of major mergers with mass ratio
$>1:3$. The \emph{dotted} and \emph{solid} lines refer to the mean
number of wet and dry mergers, defined to have a (cold) gas-to-total
mass fraction in the progenitors higher and lower than 0.15,
respectively.}\label{fig|NumberMergersFromModels}}
\end{figure*}

If the problem with the models is simply the overall normalization
of the sizes for a given \mstar, then it is interesting to study
what aspect of the models produced the flat size-age relation.
The first question which arises in this context is what these
objects looked like in the past.  The red circles in each panel
show the sizes, masses and densities (top to bottom) of these
objects at the time they formed.
The middle panels show that the mass change is larger for objects
which formed at larger redshift, as one might expect, but that this
change is most dramatic for the oldest most massive galaxies (which
have increased their mass by about a factor of ten).
Comparison with the top panels shows that the sizes increase when
the masses do, and that, except in the low mass bin, the fractional
increase in size since formation is rarely larger than that in the
mass.  As a result, the densities of low mass galaxies today have
decreased since they formed, whereas this trend is less clear for
the higher mass galaxies. %Although the densities plotted in Figure~\ref{fig|RezHistory}
%are average quantities and lack any radial information, nevertheless we note that
%the average decrease in the densities of a factor of $\sim 2-3$,
%is consistent with the decrease directly measured by \citet{Bezanson09}
%obtained by comparing the inner parts of high and low-$z$ density profiles
%of massive galaxies.

Figure~\ref{fig|compactComparison} compares the models and the data
in a format which is more like Figure~\ref{fig|ReFixedMstar}:
size versus formation time for a few bins in stellar mass (we
again here make the assumption that the age of SDSS galaxies is a good
proxy for their formation time $t_{\rm form}$).
To circumvent the problem that the scaling of SAM size with
stellar mass is wrong, we show the sizes in a fixed mass
bin normalized by the mean sizes of galaxies which formed most
recently.  For the SDSS data (open diamonds), this means that
we take the values shown by the diamonds in
Figure~\ref{fig|ReFixedMstar} and divide by the value of the left-most diamond (for the galaxies with mass above $3\times 10^{10}$\msun, we consider only galaxies with ages above $\sim 3-4$ Gyr, given that the bins
below are not statistically significant); for the SAMs (filled circles), we take the blue circles shown in Figure~\ref{fig|RezHistory} divided by the
leftmost blue circle (for consistency, for the more massive galaxies in the SAM we only select objects with a minimum age of $\sim 3-4$ Gyr).
Note how the SAMs are generally in good
agreement with the data.  They produce rather flat size-age
relations, and, at low masses, the older galaxies are about $2\times$
smaller than younger ones.  We show below that, in the models,
this happens because lower mass galaxies undergo fewer mergers.

\begin{table*}
\caption{Fractional Increase In Stellar Mass and Radius\label{tab|bhmass}.\label{tab|mergerRates}}
\begin{center}
\begin{tabular}{lrrrrrrrr}
\hline
\hline
&
\multicolumn{2}{c}{$10 < \log M_{\rm star} < 10.5$}&
&
\multicolumn{2}{c}{$10.5 < \log M_{\rm star} < 11$}&
&
\multicolumn{2}{c}{$11 < \log M_{\rm star} < 11.5$}
\\
\cline{2-3}
\cline{5-6}
\cline{8-9}
\multicolumn{1}{l}{Type} &
\multicolumn{1}{c}{$R$} &
\multicolumn{1}{c}{$M_{\rm STAR}$} &
&
\multicolumn{1}{c}{$R$} &
\multicolumn{1}{c}{$M_{\rm STAR}$} &
&
\multicolumn{1}{c}{$R$} &
\multicolumn{1}{c}{$M_{\rm STAR}$}
\\
\hline
     Minor Dry Mergers & 57\% & 33\%&   & 73\%& 30\%&    &90\% &36\% \\
     Major Dry Mergers & 26\% & 49\%&   & 22.5\%& 59\%&    &0\% &61\% \\
     Wet Growth & 10\% & 8\%&   & 4\%& 2\%&    &2.5\% &1\% \\
     Disk Instability & 7\%& 10\%&   & 0.5\%& 9\%&    &7.5\% &2\% \\
\hline
\end{tabular}
\end{center}
\raggedright Notes: Fractional increase in size and stellar mass for
 the galaxies of different mass at $z=0$; although most of the stellar
 mass is added via major dry mergers, most of the growth in size
 is through minor dry mergers.
\end{table*}

\subsection{Major versus minor mergers}
Figure~\ref{fig|NumberMergersFromModels} shows the predicted mean
number $N_{\rm MERGERS}$ of wet (dotted) and dry (solid), minor (left) or major (right)
mergers that today's early-type galaxies have undergone since they
formed. More specifically, we compute
the mean number of mergers per Gyr a galaxy had since its
formation epoch, averaged
over all galaxies, as a function of lookback time $t$.
(The numbers of mergers for each galaxy were extracted from the same 100 merger
trees used in the previous figures;
wet mergers have a cold-total gas mass fraction in the progenitors
that is greater than 0.15, else the merger is dry;
minor mergers have mass ratios $<1:3$, else the merger is major.)
Notice that there are essentially no wet mergers;
 massive objects have had an order of magnitude more major
 mergers and at least a factor of two more minor mergers, than
 lower mass objects; and
 that minor mergers are typically a factor of 5 times more frequent
 than major mergers.

Table 1 summarizes the actual increases in size and stellar mass
experienced by the early-type galaxies in the model. It is apparent
that while a substantial fraction of the stellar mass is added to
the galaxy via major mergers, the sizes mostly increase via
\emph{minor} dry mergers since their formation epoch.
These galaxies remain gas-poor for most of their assembly history:
only $\lesssim 10\%$ (decreasing to $\lesssim 4\%$ for the most massive
systems) of the final sizes and masses grow due to gas-rich mergers.
These objects are all bulge-dominated, given that $\lesssim 10\%$ of
the size and stellar mass is increased via disk instability.

The minor merger-dominated size evolution of massive early-type
galaxies in Figure~\ref{fig|NumberMergersFromModels} may be related
to that of their DM halos. \citet{Stewart08} from high-resolution
$\Lambda$CDM N-body simulations found that the mass assembly in
``galactic'' halos, those with mass in the range $10^{11}-10^{13}\,
h^{-1}\, $\msun, to be dominated by mergers that are $\sim$ 10\% of
the final halo mass.
Minor dry mergers would also more easily preserve the projections
of the fundamental plane, such as the the $L$-\sis\
\citep[e.g.,][]{FJ,Davies83} and $R_e$-$L$
\citep[e.g.,][]{Kormendy77,Ziegler99,Bernardi03} relations
(see discussion in \citealt{CiottiReview,Bernardi09} and references therein).
Preliminary measurements also show that the central densities within the
same physical scale for samples of low and high redshift galaxies of the
same stellar mass, are consistent within a factor of $\sim 2$
\citep[e.g.,][]{Cimatti08,Bezanson09}. The latter findings
could be consistent with an inside-out evolutionary scenario, where
stellar matter is continuously added to the outskirts of the compact
high-redshift galaxies as time goes on.
However, larger samples of galaxies at different redshifts with well
measured density and metallicity radial profiles are required to set
definite conclusions
\citep[e.g.,][]{Cimatti08,Bezanson09,Hopkins09CompactGalx}.
In particular, \citet[][see also Shankar \& Bernardi 2009]{Bezanson09}
discussed the results of several basic
models for the size and mass evolution of spheroids. Overall, they conclude
that galaxies with stellar mass $\sim 10^{11}$\msun\ at $z\sim 2$,
should undergo about $8$ minor mergers to efficiently increase
their sizes by a factor of $\sim 5$ and mass of $\sim 2$ to grow onto
the local size-mass relation. They also note that
their central densities would then be consistent (within a factor of $\sim 2$)
with the ones of SDSS galaxies with mass a factor of $\sim 2$ higher, further supporting
the minor merger hypothesis.
However, the cumulative number of minor mergers in the B06 model is significantly
lower than the one put forward by \citet{Bezanson09} (see also \PapI), and
in fact the size increase in the mock massive galaxies presented
here is hardly enough to bring them onto the local size-mass relation,
as discussed above. Moreover, \citet{Nipoti09} recently concluded
through a series of N-body simulations, that reproducing
the growth of a factor of about $\sim 5$ only
through dry mergers
is problematic. They find mergers not to efficiently
grow galaxies in the required proportion, and to increase the scatter in the 
galactic scaling relations beyond what allowed by observations.
They therefore conclude that  
observational biases in the measurement of the 
compact high-$z$ galaxy sizes, coupled to extreme
fine-tuning in the merger processes, 
are required to accommodate a pure merger-driven
scenario as main driver for the size evolution
of ellipticals.

\section{DISCUSSION}
\label{sec|discu}

%\section{Comparison with van der Wel et al. (2009)}
While we were completing this work, we became aware of the study
of the age-size relation by \citet{vanderwel09}.
They too find that SDSS galaxies show no relation between size and
age.  Using simple prescriptions for the merger histories of
galaxies between their formation redshift and the present, they
also conclude that models in which galaxies grow through dry mergers
are consistent with the observed evolution since $z\sim 2$ in the
mean size and in the co-moving mass density.  However, as also recently addressed
by \citet{Bernardi09b}, there is one important respect in which our results differ
from theirs.

\begin{figure}%[ht!]
\includegraphics[width=0.99\hsize]{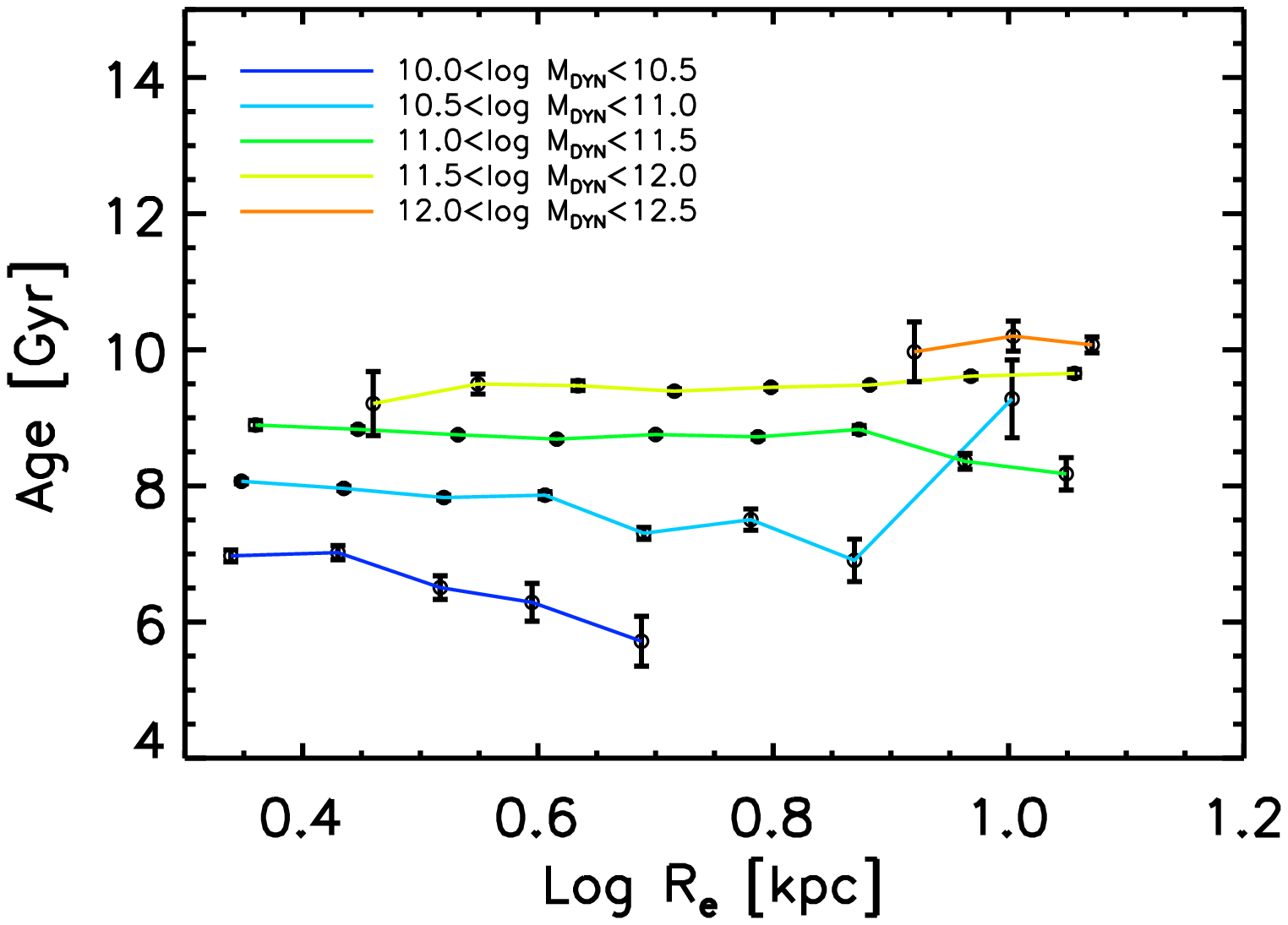}
\includegraphics[width=0.99\hsize]{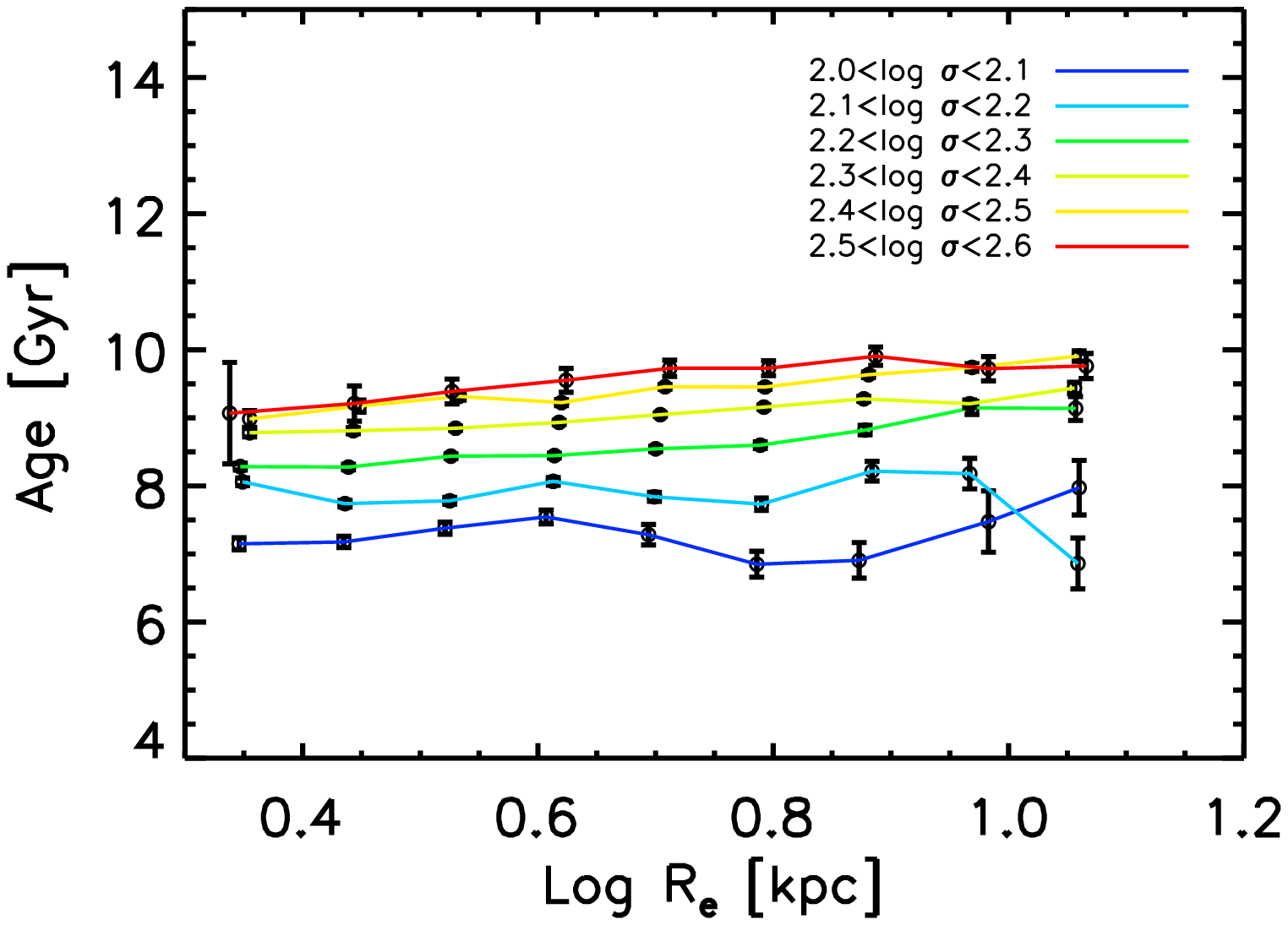}
{\caption{Age-size relations at fixed dynamical mass $M_{\rm dyn}$
(upper panels), and at fixed velocity dispersion $\sigma$ (lower panels).
}
\label{fig|comparevdWel}}
\end{figure}

Although we have studied the size as a function of age, their
Figure~1 shows the age as a function of size.  To enable comparison
with their work, our Figure~\ref{fig|comparevdWel} shows this relation
at fixed $M_{\rm dyn}$ and $\sigma$ in our dataset.  While we agree
with them that, at fixed $\sigma$, this relation is weak, we come to a
somewhat different conclusion about this relation at fixed $M_{\rm dyn}$.
Whereas they find that, at fixed $M_{\rm dyn}$, smaller galaxies are
older, we find no correlation between age and size at fixed $M_{\rm dyn}$.
This is consistent with our other results above, if one
allows for the fact that $M_{\rm STAR}$ and $M_{\rm dyn}$ are closely
related, so it is a reasonable approximation to substitute one for the
other.
\citet{Bernardi09b} show that this discrepancy is almost certainly
due to the fact that our sample is less contaminated by objects with
disks: whereas ellipticals have a flat age-size relation at fixed Mdyn,
age and size are anti-correlated for S0s and Sas.
This suggests that the two early-type galaxy populations have different
formation histories.

%\subsection{Comparison with Besanzon et al. (2009)}

\section{CONCLUSIONS}\label{sec|conclu}

We studied the size-mass-age relations as derived from a sample of
about 45,700 early-type galaxies selected from SDSS,
and compared our results with Bower et al.'s (2006) model of
hierarchical galaxy formation.
Our results can be summarized as follows.
\begin{itemize}
\item
At stellar masses below $3\times 10^{10}\, $\msun,
the effective radius \re\ can be up to a factor of $\sim 2$ lower
for older galaxies.  However, more massive galaxies all share
similar size distributions, irrespective of their ages
(Figure~\ref{fig|ReFixedMstar}).
At these higher masses, the scatter in sizes at fixed stellar mass
is about a factor of~4 ($\sim 0.6$ dex), decreasing
with increasing stellar mass (Figure~\ref{fig|RevsMstarGlobal}).
These findings are at variance with a pure passive evolution, which
would predict older galaxies to be much more compact at fixed stellar mass.
\item Hierarchical galaxy formation models predict that galaxies
of the same stellar mass formed at high redshifts are much more
compact and dense (Figure~\ref{fig|RezModelsFixedMstar}), in agreement
with observations at higher redshift (see \S~\ref{sec|intro}).
This is both because the high redshift Universe itself is denser,
and because dissipation effects are more effective at early times.
\item SAMs based on a hierarchical growth of galaxies driven by a
first major, wet merger and a sequence of late, minor, dry mergers,
predict that these extremely small, high-redshift galaxies
can grow, on average, onto the same local size-age relation
(Figures~\ref{fig|RezHistory} and \ref{fig|compactComparison}).
Note, however, that (dry) mergers are not the unique way to increase
early-type galaxy sizes.  For example, \citet{Fan08}, put forward a
model that postulates a strong galaxy expansion caused by the mass
loss due to quasar feedback and stellar winds.
This model predicts a local size-age relation that is consistent with
that one observed in SDSS, at least at lower masses (see further
discussion in \citealt{ShankarBernardi}) -- something which the
SAMs are unable to accomplish.
\item In the SAMs, galaxies which form at higher redshifts experience
more mergers than galaxies which formed more recently, increasing their
original sizes by a greater factor than galaxies which formed later.
This process almost completely wipes out the monolithic effect,
growing all galaxies towards the same size-mass relation today.
In particular, although most of the stellar mass is added via
major dry mergers, most of the growth in size is through minor dry mergers.
Minor mergers (mass ratios $<1:3$) outnumber major mergers by about
a factor of 5 at \mstar$>10^{10.5}M_\odot$, and by a factor of 10 at
smaller masses (Figure~\ref{fig|NumberMergersFromModels} and
Table~\ref{tab|mergerRates}).
\item However, these SAMs provide a poor match to the local size-mass
relation, and much more work has to be done
to understand the origin of these discrepancies (e.g. \citealt{Trujillo09},
\citealt{Taylor09}, and \PapInp).
\end{itemize}

\section*{Acknowledgments}
We thank Guinevere Kauffmann,
Mike Boylan-Kolchin, Philip Hopkins, David Wake, and Tommaso Treu
for interesting discussions. We also thank the referee for several
useful comments and suggestions that improved the presentation
of the paper. FS acknowledges support from the Alexander von Humboldt Foundation
and NASA Grant NNG05GH77G. MB is supported by NASA grant
LTSA-NNG06GC19G and NASA ADP/NNX09AD02G.
MB and RKS thank the Observatory
at Meudon, the APC, Paris-7, and MPI-Astronomie Heidelberg,
for hospitality when this work was completed.

\bibliographystyle{mn2e}
\bibliography{../../RefMajor}

%\appendix

\label{lastpage}

\end{document}